\documentclass[journal]{IEEEtran}
\usepackage{nopageno}
\usepackage{float}
\usepackage{caption}
\usepackage{subfig}

\ifCLASSINFOpdf
   \usepackage[pdftex]{graphicx}
   \graphicspath{{images/}}
\else
\fi

\usepackage{amsmath}
\usepackage{xcolor}
\begin{document}
\title{Direct Iterative Reconstruction of Multiple Basis Material Images in Photon-counting Spectral CT}
\author{Obaidullah~Rahman,
        Ken~D.~Sauer,~Department~of~Electrical~Engineering,~University~of~Notre~Dame\\
        Connor~J.~Evans,~Ryan~K.~Roeder,~Department~of~Aerospace~and~Mechanical~Engineering,~University~of~Notre~Dame}

\maketitle
\begin{abstract}
In this work, we perform direct material reconstruction from spectral CT data using a model based iterative reconstruction  (MBIR) approach. 
Material concentrations are measured in volume fractions, whose total
is constrained by a maximum of unity. 
A phantom containing a combination of 4 basis materials (water, iodine, gadolinium, calcium) was scanned using a photon-counting detector. 
Iodine and gadolinium were chosen because of their common use as contrast agents in CT imaging. 
Scan data was binned into 5 energy (keV) levels.
Each energy bin in a calibration scan was reconstructed, allowing the linear attenuation coefficient of each material for every energy to be estimated by a 
least-squares fit to ground truth in the image domain.
The resulting $5\times 4$ matrix, for $5$ energies and $4$ materials,
is incorporated into the forward model in direct reconstruction of the $4$ basis material images with
spatial and/or inter-material regularization.
In reconstruction from a subsequent low-concentration scan, volume fractions within regions of interest (ROIs) are found to be close to the ground truth. 
This work is meant to lay the foundation for further work with phantoms including spatially coincident mixtures of contrast materials and/or contrast agents in widely varying concentrations, molecular imaging from animal scans, and eventually clinical applications.
\end{abstract}


\IEEEpeerreviewmaketitle

\section{Introduction}
\IEEEPARstart{S}{pectral} CT has shown great promise for a variety of medical and scientific imaging tasks.
The technology is enabled by the advent of photon-counting detectors (PCD), which
can accurately count the number of received photons and bin them according to energy levels \cite{Roessl}.  
Energy bins can be adjusted to leverage the K-edge of some contrast agents for better delineation among basis materials. 
This also allows us to more closely model the attenuation effects of materials at differing energy levels, limiting the effects of beam hardening artifacts. 
This permits a more accurate linear model between image and sinogram domains.

Material decomposition is typically accomplished by reconstructing each energy
bin independently as a monochromatic scan using a traditional algorithm such as 
FBP or an iterative technique such as model-based iterative reconstruction (MBIR). 
The resulting reconstructions are linear attenuation coefficient (LAC) images for
respective energy levels which are then, through a ``mixing matrix" describing linear attenuation by
each material at each energy level, decomposed into basis material fraction images\cite{Curtis2}. 
For greater accuracy than with material modeling from standard values, the mixing matrix can be calibrated using phantoms
with known materials and concentrations\cite{Curtis}.
Calibration is meant to account for system specific variations. 
In this work, we use a calibration dataset to compute our mixing matrix, 
$\mathcal{M}$ from individual LAC images of $5$ bins and the ground truth. 
Then we use that $\mathcal{M}$ in the linear model to reconstruct material fraction images directly from subsequent scans
containing similar materials.

The goal of this work is to detect, identify and quantify contrast agents within a phantom using MBIR directly 
from spectral CT sinogram data, and to lay a foundation for its eventual use in laboratory and clinical studies. 
Water, a natural choice for one of the basis materials, was chosen to simulate soft tissue. 
Iodine and gadolinium were chosen because of their wide usage as contrast agents in diagnostic imaging. 
Calcium was chosen to simulate mineralized tissue. 
We are attempting to estimate material fractions the order of $10^{-3}$ for contrast agents. 
The challenges in this work come from low signal contributions from the  contrast which,
with relatively large condition numbers of mixing matrices,
poses a challenging estimation problem as the solutions become more dilute. 

A great deal of research has been published in the area of spectral CT, a large fraction of it dealing
with dual-energy systems \cite{Zhang,Mendonça}.
Direct reconstructions to date have typically used  fewer energy bins \cite{Long} or fewer basis materials than our current efforts.
Some have worked on more than 2 energies \cite{Schmidt} but with $2$ basis materials only. 
Our previous work has been focused on inversion from multi-energy attenuation coefficients
to material concentrations under constrained optimization in the image domain\cite{Curtis2}.
The present work departs from previous in that we attempt direct reconstructions at low contrast among
multiple contrast agents using a direct statistical iterative reconstruction algorithm in spectral CT. 

\section{Method}
\subsection{Background}
\subsubsection{MAP estimation for poly-chromatic case}
Let $x = [x_l ; l\in (1,L)]$ be the vector in image space.
In conventional, monochromatic CT, $x$ will represent
linear attenuation coefficients for a single scan, while in
our subsequent spectral case, it will represent concentrations
of various materials in an expanded form.
The vector $y = [y_n ; n\in (1,N)]$ contains projection
measurements.
Let $A_{N\times L}$ be the projection matrix that maps linear attenuation to projection space
according to scanner geometry. 
In the work presented here, $A$ is computed according to the
methods described in \cite{Thibault}.
The MAP estimate of $x$ for a single energy can be formulated by
\begin{eqnarray}
\hat{x} = \underset{x}{\operatorname{\textit{argmin}}} \ p(y|x)g(x),
\end{eqnarray}
with $p$ and $g$ representing conditional probability density of
measurements and an {\it a priori} model of the image, respectively.
The MAP estimate of $x$ \cite{Bouman} can be approximated as
\begin{eqnarray}
\hat{x} = \underset{x}{\operatorname{\textit{argmin}}} \ \frac{1}{2}(y-Ax)^TD(y-Ax) + U(x)
\label{eq:approxMAP}
\end{eqnarray}
where $D$ is a weighting matrix approximately inverse to conditional variance of data
and $U(x)$ is a log-prior on the image. 
With an iterative coordinate descent (ICD) scalar update, a $k-th$ step in optimization of (\ref{eq:approxMAP})
becomes
\begin{eqnarray}
\hat{\Delta x_l^k} = \underset{\Delta x_l}{\operatorname{\textit{argmin}}} \frac{1}{2}(e^k+A_{*l}\Delta x_l)^TD(e^k+A_{*l}\Delta x_l) \nonumber \\
+ U_l(x+I_l\Delta x_l)
\label{eq:ICDupdate}
\end{eqnarray}
where $\Delta x_l^k$ is the change in image value $x_l$ at update $k$,
$e^k = y-Ax^k$, the sinogram error vector at step $k$,
$A_{*l}$ represents the $l^{th}$ column of $A$,
$I_l$ is zero except for unity at site $l$,
and $x_l^{k+1} = x_l^k+ \Delta x_l^k$.
ICD may have advantages in convergence, and in our case,
simpler computation under varied image models for $U(x)$ is possible.

\subsubsection{MAP estimation for spectral case}
MAP estimation for the polychromatic case can be easily extended to spectral data.
 In this case, $x = [x_l ; l\in (1,L)]$ will again be the vector of image space with $L$ voxels. 
However, each entry $x_l$ is now an $M$-element vector such that $x_l = [x_{lm}; m \in (1,M)]$, 
where $M$ is the number of basis materials. 
$x_{L\times M}$ will represent the {\it volume fraction} for each material, rather than linear attenuation. 
$y = [y_e ; e \in (1,E)]$, where $E$ is the number of energy bins. 
$y_e$ represents $e^{th}$ energy sinogram. 
Each entry in $y_e$ is a $N$-element vector. $y_e = [y_{en}; n\in (1,N]$, where $N$ is the number of sinogram datapoints in each energy bin.
 $y_{E\times N}$ represents total sinogram measurement. 
 Let $A_{N\times L}$ be the standard projection matrix as before. 
 $x$ can be estimated by the following equation:
\begin{eqnarray}
\hat{x} = \underset{x}{\operatorname{\textit{argmin}}} \bigg{[}\frac{1}{2} \sum_{e=1}^{E} (y_e- A (\mathcal{M}x^T)^T)^T D (y_e- A (\mathcal{M}x^T)^T)\nonumber\\
+ U(x)\bigg{]}
\label{eq:spectral_apost}
\end{eqnarray}
The dimension of $y_e$ is $N \times 1$, $\mathcal{M}$ is $E \times M$, $X^T$ is $M \times L$, $\mathcal{M}X^T$ is $E \times L$, $D$ is $N\times N$, and $A (\mathcal{M}x^T)^T$ is $N \times 1$ which is same as $y_e$.

The vector ICD update for each voxel is identical to (\ref{eq:ICDupdate}),
except that $x_l$ is a vector which is converted from material fractions to 
linear attenuation by $\mathcal{M}$ for forward projection by $A$ before insertion into the log-likelihood term.

\subsection{ Optimization }
\subsubsection{Convexity}
The Hessian of first cost term w.r.t. $x_l$ is positive definite, since it is quadratic in $x_l$ with a
positive, diagonal weighting. 
We apply a non-negativity constraint on all material fraction values, 
and force the material fractions for a voxel to sum to no more than $1$. 
The optimization problem now becomes the modified equation (\ref{eq:ICDupdate}) with addition of the following constraints
\begin{eqnarray}
\text{s.t., }x_l \geq \mathbf{0} \\
x_l^T \mathbf{1} \leq 1 \nonumber
\end{eqnarray}
where $\mathbf{1} \text{ and } \mathbf{0}$ are $M$-element vectors with all ones and all zeros respectively. Since the cost is convex and the constraints are linear, it is a convex problem with a unique minimum. 
During every ICD voxel update, the M-dimensional optimization problem is solved. 
We employ a modified simplex method to solve the problem.

The regularizer term $U(x_l)$ was chosen to be a simple quadratic function that penalizes the difference in the current voxel and its neighbors.
\begin{eqnarray}
U(x_l) = \frac{1}{2\sigma^2}\sum_{m=1}^{M} \sum_{k\in\mathcal{N}_l} \alpha_k (x_{lm} - x_{km})^2
\end{eqnarray}
where $\mathcal{N}_l$ denotes the neighborhood of $x_l$, and $\alpha_k$ depends on the position of the neighbor such that $\sum_{k\in \mathcal{N}_l} \alpha_k = 1$. 
Its Hessian w.r.t. $x_l$ is a diagonal matrix with all entries $\frac{1}{\sigma^2}$, since
we have not yet applied inter-material regularization. 
The Hessian of total cost stays positive definite and it still has a unique minimum.
If we are concerned with the average value of material fraction within ROIs, we can afford to have somewhat noisy reconstructions,
as this helps avoid bias. 
Therefore the value of $\sigma$ was chosen to yield very light regularization for current experiments.

\subsubsection{Simplex Implementation}
Since both terms in cost are quadratic, it can be re-written in a standard quadratic expression. 
Let $\mathcal{H}$ be the hessian
and $\mathcal{D}$ be the derivative of the total cost w.r.t $x_l$. 
The optimization problem can be written in vector notation as
\begin{eqnarray}
\frac{1}{2} x_l^T\mathcal{H}x_l + \mathcal{D}^Tx_l
\end{eqnarray}
The single-site optimization problem, from the KKT condition, now becomes
$$\hat{x_l} = \underset{x_l}{\operatorname{\textit{argmin}}} \ \frac{1}{2} x_l^T\mathcal{H}x_l + \mathcal{D}^Tx_l - \mathbf{\mu}^Tx_l + \lambda (\mathbf{1}^T x_l - 1)$$
\begin{align}
&\text{s.t., } \mathbf{1}^T x_l \leq 1 \\
x_l \geq \mathbf{0},\ \mathbf{\mu} \geq \mathbf{0},\ \lambda &\geq 0,\ \mu_m x_{lm} = 0 \text{ for } m = 1 \text{ to } M \nonumber
\end{align}
where $\mathbf{\mu}$ is a vector Lagrangian multiplier for positivity constraint, $\lambda$ is a scalar Lagrangian
for the sum-less-than-or-equal-to-$1$ constraint. 
$\mu_m x_{lm} = 0 $ comes from complementary slackness. 
We solve this problem during every ICD voxel update using a modified simplex method.

\subsection{Spectral CT Data Acquisition}
Phantom solutions were created with varying concentrations of iodine, gadolinium, and calcium in water in 7 tubes. 
Please refer to Table \ref{tab:Table1} for concentration and spatial information of different solutions.
Note that iodine, gadolinium, and calcium are not mixed with each other and are kept spatially orthogonal in this setup. 
We intend to mix contrasts in our future work.
$K$-edges of iodine and gadolinium are around $33$ KeV and $53$ KeV respectively. 
The bins are chosen to take advantage of the difference in the photon absorption properties of different basis materials.
Some defective detector pixels did not register any signal, or were unreliable, and
data correction was therefore performed to eliminate outliers and make sinograms consistent.
The first set of phantom solutions was used to calibrate the mixing matrix (please refer to the $2^{nd}$ column of Table \ref{tab:Table1}). 
Then that mixing matrix was used in the forward model of reconstruction for a $2^{nd}$ set of phantom data  (please refer to the $3^{rd}$ column of Table \ref{tab:Table1}).

\begin{table}
	\begin{minipage}{0.5\linewidth}
		\begin{tabular}{|c|c|c|c|c|}
			\hline
			Material & Calibration  & Test\\
			\hline
			& $Conc. $ & $Conc.$\\
			\hline
			\textcolor{red}{$I_{high}$} & \textcolor{red}{15.86} & \textcolor{red}{9.52}\\
			\hline
			\textcolor{red}{$I_{low}$} & \textcolor{red}{7.93} & \textcolor{red}{4.76}\\
			\hline
			\textcolor{green}{$Gd_{high}$} & \textcolor{green}{19.66} & \textcolor{green}{11.79}\\
			\hline
			\textcolor{green}{$Gd_{low}$} & \textcolor{green}{9.83} & \textcolor{green}{5.90}\\
			\hline
			\textcolor{magenta}{$Ca_{high}$} & \textcolor{magenta}{146.29} & \textcolor{magenta}{87.77}\\
			\hline
			\textcolor{magenta}{$Ca_{low}$} & \textcolor{magenta}{73.14} & \textcolor{magenta}{43.89}\\
			\hline
		\end{tabular}
		\caption{\label{tab:Table1} \footnotesize Concentration of materials for Calibration data and Test data in $mg/mL$}
	\end{minipage}\hfill
	\begin{minipage}{0.45\linewidth}
		\centering
		{\includegraphics[width = 1.1in]{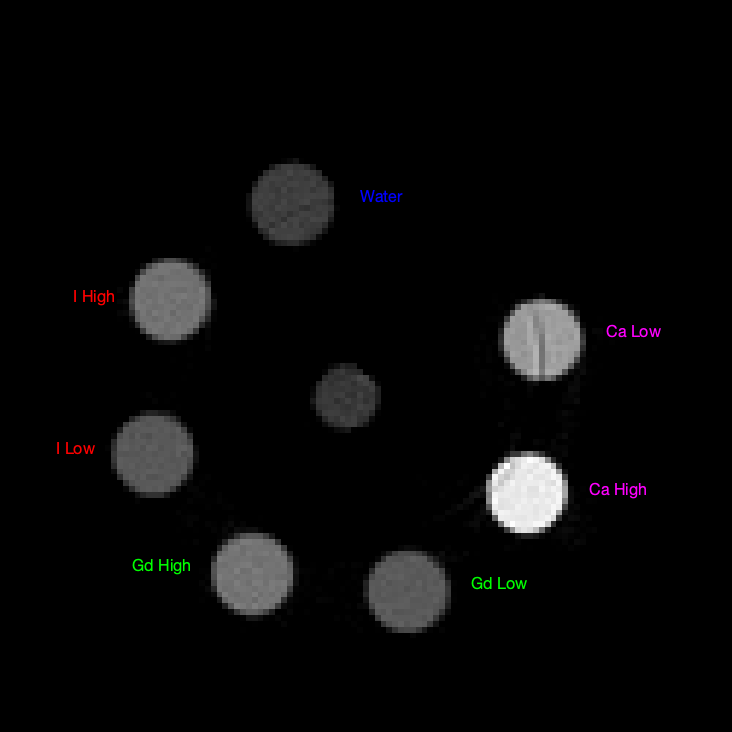}\label{fig:Material_label}}
		\captionof{figure}{{\footnotesize Material Concentration and spatial location}}
	\end{minipage}
\end{table}

The scanner employed is a photon-counting spectral CT scanner manufactured by MARS Bioimaging (Christchurch, NZ). 
The photon-counting detector comprises 5 CdZnTe sensors separated by small gaps and bonded to a Medipix 3RX chip.
Each detector panel has $128$ rows and $128$ channels, and $0.11\ mm\times 0.11\ mm$ ``pixel pitch". 
Scan voltage was $80\ kVp$ with an X-ray tube current of $35\ \mu A$.
The total number of views was $720$ in an axial scan mode.
The resulting dimension of the sinogram was $128$ rows, $663$ channels, $740$ views. 

Source to iso-center distance was $200\ mm$, and source to detector distance was $250\ mm$. 
The energy bins are (1) $7.0-19.0$, (2) $19.0-29.0$, (3) $29.0-38.8$, (4) $38.8-51.1$, and (5) $51.1-82.6 \ keV$.
The photon count rate per pixel was kept to 10 photons/ms to mitigate pulse pileup effects.
The exposure time was 125 ms.
No external beam filtration was used.
Sensor pixels collect with "charge-summing mode."
Average counts per detector pixel was $1500$.
Voxel size of reconstructions was $0.4\ mm \times 0.4\ mm \times 0.5\ mm$.
Reconstruction field of view was $61\ mm$.

\subsection{Mixing matrix computation}
\begin{figure}[h]
\centering
\subfloat[][]{\includegraphics[width = 1.0in]{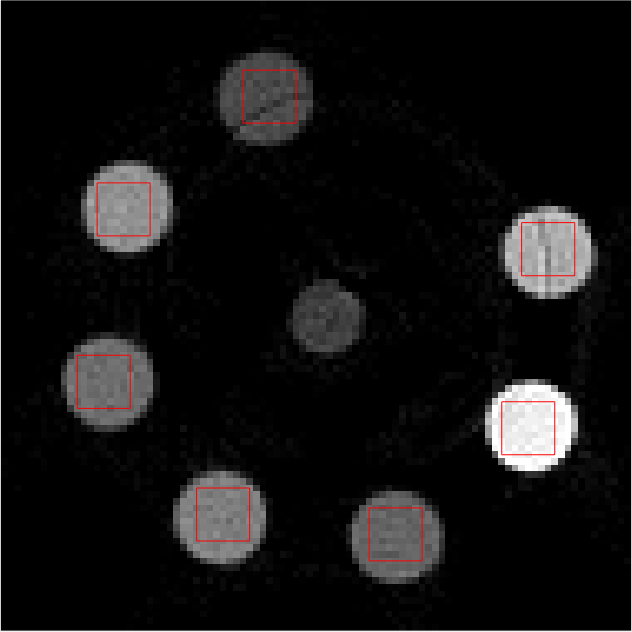}\label{fig:En1}}
\hspace{0.1 cm}
\subfloat[][]{\includegraphics[width = 1.0in]{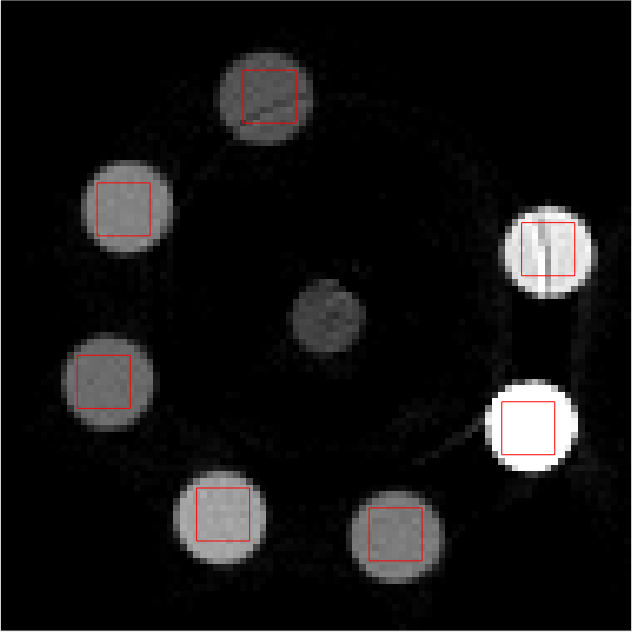}\label{fig:En2}}
\hspace{0.1 cm}
\subfloat[][]{\includegraphics[width = 1.0in]{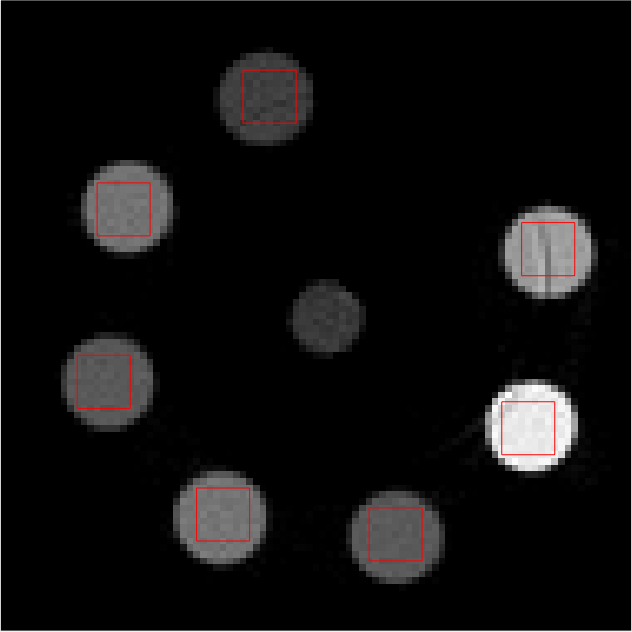}\label{fig:En3}}
\hspace{0.1 cm}
\subfloat[][]{\includegraphics[width = 1.0in]{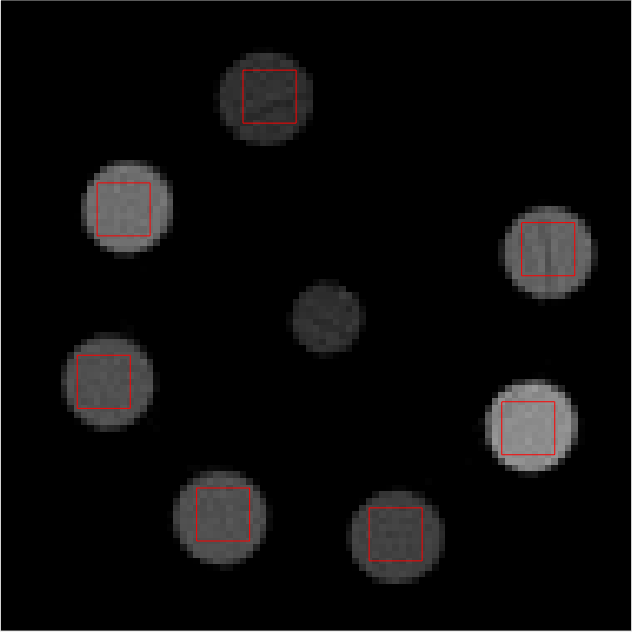}\label{fig:En4}}
\hspace{0.1 cm}
\subfloat[][]{\includegraphics[width = 1.0in]{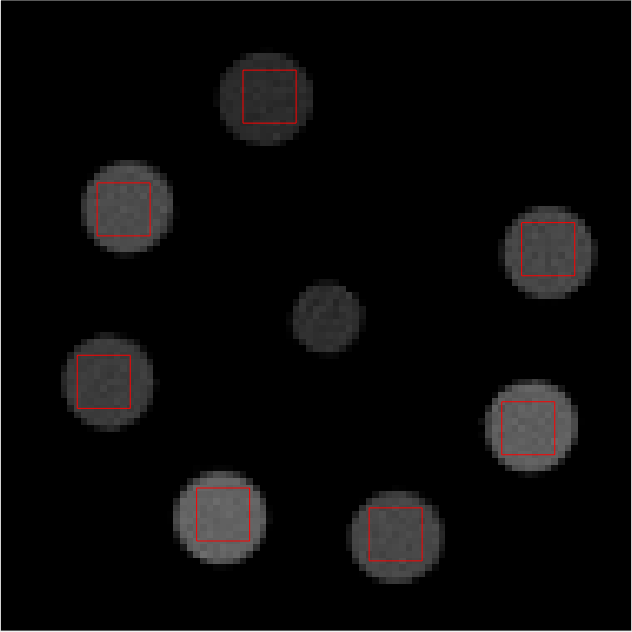}\label{fig:En5}}
\caption{{\footnotesize Middle slice of LAC image reconstruction of each energy bin sinogram for mixing matrix
estimation. 
Display window [0, 0.1]}\label{fig:figure1}}{\footnotesize \subref{fig:En1} Bin1:$7.0-19.0 \ keV$, \subref{fig:En2} Bin2: $19.0-29.0 \ keV$, \subref{fig:En3} Bin3: $29.0-38.8 \ keV$, \subref{fig:En4} Bin4: $38.8-51.1 \ keV$, \subref{fig:En5} Bin5: $51.1-82.6 \ keV$}
\end{figure}

Individual reconstructions from every bin were first performed with scalar MBIR.
This furnished $5$ LAC images, each from a different energy bin. 
$10$ slices of each LAC image volume were used to compute $\mathcal{M}$.
The mixing matrix was computed from ROIs shown in red in Fig \ref{fig:figure1}.
The following equation must be satisfied ideally
\begin{eqnarray}
\mathcal{M}x^T = \mu
\end{eqnarray}
where $x$ is a $Q\times M$ matrix that represents material fraction of $Q$ voxels in 
ROIs that is known from ground truth, $\mu$ is a $E\times Q$ matrix that denotes LAC of $Q$ voxel in ROIs from $E$ reconstructed bins. 
$\mathcal{M}$ is computed with least-squares linear regression using the following
\begin{eqnarray}
\mathcal{M} = \mu x(x^Tx)^{-1}
\end{eqnarray}
The resulting attenuation coefficients may be found in Table \ref{tab:Table2}.

\begin{table}
\begin{center}
\begin{tabular}{|c|c|c|c|c|}
\hline
Energy & \multicolumn{4}{c}{Linear Attenuation Coefficient}\\
\hline
& $Water$ & $Iodine$ & $Gadolinium$ & $Calcium$\\
\hline
$Bin\ 1$ & 0.0301 & 8.0544 & 7.5169 & 0.7722\\
\hline
$Bin\ 2$ & 0.0345 & 4.9786 & 11.1868 & 1.0905\\
\hline
$Bin\ 3$ & 0.0253 & 5.9366 & 8.1421 & 0.7815\\
\hline
$Bin\ 4$ & 0.0196 & 7.2125& 4.8177 & 0.4259\\
\hline
$Bin\ 5$ & 0.0177 & 3.7628 & 8.4091 & 0.2395\\
\hline
\end{tabular}
\caption{\label{tab:Table2} \footnotesize Mixing matrix computed from 5 sinograms and ground truth. All values of linear attenuation coefficient are in $mm^{-1}$}
\end{center}
\end{table}

\subsection{Spectral Reconstruction}
MBIR of basis material fractional concentration images is now performed directly from the collection of 5 new sinograms.
The output reconstruction images are material maps for the basis materials' volume fractions. 
The result presented here comes from iteration $150$.

\section{Results}
The computed mixing matrix $\mathcal{M}$ is used in the model based iterative algorithm to reconstruct the material fraction images. 
The convergence is slow because of the poor condition number of the Hessian in (\ref{eq:spectral_apost}). 
Since the tube solutions are nearly constant in $z$ direction, we show only one reconstructed slice of the total image volume. 
Each ROI is the center square patch of the 2-D image within each tube.

\begin{figure}[h]
\centering
\subfloat[][]{\includegraphics[width = 1.5in]{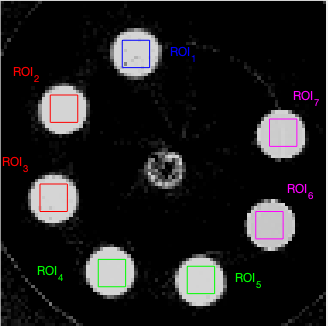}\label{fig:Water}}
\hspace{0.1 cm}
\subfloat[][]{\includegraphics[width = 1.5in]{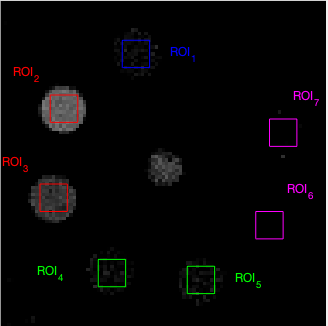}\label{fig:Iodine}}
\hspace{0.1 cm}
\subfloat[][]{\includegraphics[width = 1.5in]{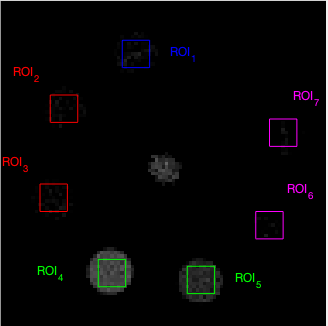}\label{fig:Gadolinium}}
\hspace{0.1 cm}
\subfloat[][]{\includegraphics[width = 1.5in]{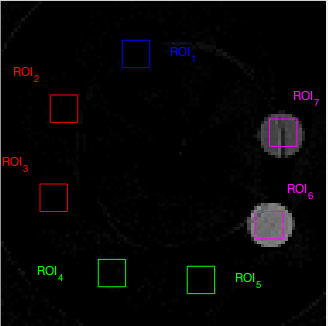}\label{fig:Calcium}}
\caption{{\footnotesize Reconstructed image of material fraction of }\label{fig:figure2}}{\footnotesize \subref{fig:Water} Water: Display window [0, 1.2]. \subref{fig:Iodine} Iodine: Display window [0, 0.006]. \subref{fig:Gadolinium} Gadolinium: Display window [0, 0.006]. \subref{fig:Calcium} Calcium: Display window [0, 0.15]}
\end{figure}

\begin{table}[!h]
\begin{center}
\begin{tabular}{|c|c|c|c|c|}
\hline
\multicolumn{5}{c}{Ground Truth (volume fraction)}\\
\hline
Spatial Loc. & \multicolumn{4}{c}{Basis Material}\\
\hline
& $Water$ & $Iodine$ & $Gadolinium$ & $Calcium$\\
\hline
\textcolor{blue}{$ROI_1$} & \textcolor{blue}{\textbf{1.000000}} & \textcolor{blue}{0} & \textcolor{blue}{0} & \textcolor{blue}{0}\\
\hline
\textcolor{red}{$ROI_2$} & \textcolor{red}{\textbf{0.998073}} & \textcolor{red}{\textbf{0.001926}} & \textcolor{red}{0} & \textcolor{red}{0}\\
\hline
\textcolor{red}{$ROI_3$}  & \textcolor{red}{\textbf{0.999036}} & \textcolor{red}{\textbf{0.000964}} & \textcolor{red}{0} & \textcolor{red}{0}\\
\hline
\textcolor{green}{$ROI_4$}  & \textcolor{green}{\textbf{0.998509}} & \textcolor{green}{0} & \textcolor{green}{\textbf{0.001490}} & \textcolor{green}{0}\\
\hline
\textcolor{green}{$ROI_5$}  & \textcolor{green}{\textbf{0.999253}} & \textcolor{green}{0} & \textcolor{green}{\textbf{0.000746}} & \textcolor{green}{0}\\
\hline
\textcolor{magenta}{$ROI_6$}  & \textcolor{magenta}{\textbf{0.946408}} & \textcolor{magenta}{0} & \textcolor{magenta}{0} & \textcolor{magenta}{\textbf{0.053591}}\\
\hline
\textcolor{magenta}{$ROI_7$}  & \textcolor{magenta}{\textbf{0.972463}} & \textcolor{magenta}{0} & \textcolor{magenta}{0} & \textcolor{magenta}{\textbf{0.027536}}\\
\hline
\end{tabular}
\caption{\label{tab:Table3} \footnotesize Ground Truth from the phantom solutions. Contrast agents are spatially orthogonal. Note that the values in volume fraction have been converted from concentration ($mg/mL$) in Table \ref{tab:Table1} $3^{rd}$ column}
\end{center}
\end{table}

\begin{table}[!h]
\begin{center}
\begin{tabular}{|c|c|c|c|c|}
\hline
\multicolumn{5}{c}{Estimated values (volume fraction)}\\
\hline
Spatial Loc. & \multicolumn{4}{c}{Basis Material}\\
\hline
& $Water$ & $Iodine$ & $Gadolinium$ & $Calcium$\\
\hline
\textcolor{blue}{$ROI_1$} & \textcolor{blue}{\textbf{0.987255}} & \textcolor{blue}{0.000248} & \textcolor{blue}{0.000196} & \textcolor{blue}{0.000035}\\
\hline
\textcolor{red}{$ROI_2$} & \textcolor{red}{\textbf{0.997248}} & \textcolor{red}{\textbf{0.002288}} & \textcolor{red}{0.000137} & \textcolor{red}{0.000102}\\
\hline
\textcolor{red}{$ROI_3$}  & \textcolor{red}{\textbf{0.997725}} & \textcolor{red}{\textbf{0.001237}} & \textcolor{red}{0.000149} & \textcolor{red}{0.000000}\\
\hline
\textcolor{green}{$ROI_4$} & \textcolor{green}{\textbf{0.997719}} & \textcolor{green}{0.000190} & \textcolor{green}{\textbf{0.001621}} & \textcolor{green}{0.000468}\\
\hline
\textcolor{green}{$ROI_5$}  & \textcolor{green}{\textbf{0.998700}} & \textcolor{green}{0.000238} & \textcolor{green}{\textbf{0.000951}} & \textcolor{green}{0.000110}\\
\hline
\textcolor{magenta}{$ROI_6$}  & \textcolor{magenta}{\textbf{0.935048}} & \textcolor{magenta}{0.000016} & \textcolor{magenta}{0.000119} & \textcolor{magenta}{\textbf{0.064815}}\\
\hline
\textcolor{magenta}{$ROI_7$}  & \textcolor{magenta}{\textbf{0.961056}} & \textcolor{magenta}{0.000049} & \textcolor{magenta}{0.000096} & \textcolor{magenta}{\textbf{0.038797}}\\
\hline
\end{tabular}
\caption{\label{tab:Table4} \footnotesize Estimated average values of material fraction within ROIs.}
\end{center}
\end{table}

\begin{table}[!h]
\begin{center}
\begin{tabular}{|c|c|c|c|c|}
\hline
\multicolumn{5}{c}{Error (volume fraction)}\\
\hline
Spatial Loc. & \multicolumn{4}{c}{Basis Material}\\
\hline
& $Water$ & $Iodine$ & $Gadolinium$ & $Calcium$\\
\hline
\textcolor{blue}{$ROI_1$} & \textcolor{blue}{\textbf{1.27 \%}} & \textcolor{blue}{---} & \textcolor{blue}{---} & \textcolor{blue}{---}\\
\hline
\textcolor{red}{$ROI_2$} & \textcolor{red}{\textbf{0.08 \%}} & \textcolor{red}{\textbf{18.81\%}} & \textcolor{red}{---} & \textcolor{red}{---}\\
\hline
\textcolor{red}{$ROI_3$}  & \textcolor{red}{\textbf{0.13 \%}} & \textcolor{red}{\textbf{28.35\%}} & \textcolor{red}{---} & \textcolor{red}{---}\\
\hline
\textcolor{green}{$ROI_4$} & \textcolor{green}{\textbf{0.08 \%}} & \textcolor{green}{---} & \textcolor{green}{\textbf{8.80\%}} & \textcolor{green}{---}\\
\hline
\textcolor{green}{$ROI_5$}  & \textcolor{green}{\textbf{0.05 \%}} & \textcolor{green}{---} & \textcolor{green}{\textbf{27.48\%}} & \textcolor{green}{---}\\
\hline
\textcolor{magenta}{$ROI_6$}  & \textcolor{magenta}{\textbf{1.20 \%}} & \textcolor{magenta}{---} & \textcolor{magenta}{---} & \textcolor{magenta}{\textbf{20.94\%}}\\
\hline
\textcolor{magenta}{$ROI_7$}  & \textcolor{magenta}{\textbf{1.17 \%}} & \textcolor{magenta}{---} & \textcolor{magenta}{---} & \textcolor{magenta}{\textbf{40.89 \%}}\\
\hline
\end{tabular}
\caption{\label{tab:Table5} \footnotesize Error in estimation of volume fraction within ROIs (in $\%$).}
\end{center}
\end{table}

In Table \ref{tab:Table4},
We see that some amount of iodine, gadolinium, and calcium erroneously appear in some tubes,
but overall the values are close to the ground truth values of Table \ref{tab:Table3}. 
In the center of the image there is a Teflon screw that does not match any of the basis materials. 
It can be seen in the material fraction images that the outer periphery of the screw is approximated by water 
and the inside is approximated by iodine and gadolinium (and air).
As concentration decreases, in Table \ref{tab:Table5} the percentage error is observed to increase for each material, which makes sense since SNR value drops and the condition number of the mixing matrix is large.

Some of the individual reconstructed LAC images have ring artifacts and some have circular bands that have 
voxels with clearly different LAC value from their neighbors. 
This is caused by inconsistent behavior of some detectors
and contributes to error in the estimate of  material fraction's average values within the ROI. 
Another potential source of error could be the noise in mixing matrix from the calibration dataset as pointed out in \cite{Curtis}. 
Inter-material difference in material fraction values was not penalized in the regularizer term $U(x_l)$ but will be explored in future work.

\section{Conclusion}
Accurate estimation of the spectral mixing matrix is important for a good linear model. 
After computation of the matrix from a set of calibration measurements, 
direct reconstruction of material fraction images was performed successfully. 
Average values within ROIs are consistent with the ground truth.
Future work will include mixed or spatially coincident contrast agent and tissue compositions of varying concentration, as well as small-animal contrast studies. 
\section{Acknowldegement}
This work was supported by
the Martell Family PhD Fellowship and the Equipment Renewal and Restoration Program, both
at the University of Notre Dame, 
the Department of Homeland Security ALERT Center for Excellence under
Grant Award 2013-ST-061-ED0001,
the Kelly Cares Foundation and St.\ Joseph Health System.
The authors acknowledge also the assistance of Notre Dame's Integrated Imaging Facility
and Center for Environmental Science and Technology.

\end{document}